\documentclass[10pt,journal,compsoc]{IEEEtran}
\ifCLASSOPTIONcompsoc
  \usepackage[nocompress]{cite}
\else
  \usepackage{cite}
\fi
\ifCLASSINFOpdf
   \usepackage[pdftex]{graphicx}
   \DeclareGraphicsExtensions{.pdf,.jpeg,.png}
\else
\fi
\usepackage{amsmath,stmaryrd,amssymb}
\usepackage{algorithmic}
\usepackage{caption}
\usepackage{url}
\begin{document}
\onecolumn
\title{Bottleneck Time Minimization for Distributed Iterative Processes: Speeding Up Gossip-Based Federated Learning on Networked Computers}

\author{Mehrdad~Kiamari
        and~Bhaskar~Krishnamachari
        % <-this % stops a space
\IEEEcompsocitemizethanks{\IEEEcompsocthanksitem The authors are with with the Department
of Electrical and Computer Engineering, University of Southern California, Los Angeles,
CA, 90089.\protect\\
E-mail: \{kiamari,bkrishna\}@usc.edu 
}% <-this % stops an unwanted space
}


\markboth{}
{Shell \MakeLowercase{\textit{et al.}}: Bare Demo of IEEEtran.cls for Computer Society Journals}



\IEEEtitleabstractindextext{%
\begin{abstract}
We present a novel task scheduling scheme for accelerating computational applications involving distributed iterative processes that are executed on networked computing resources. Such an application consists of multiple tasks, each of which outputs data at each iteration to be processed by neighboring tasks; these dependencies between the tasks can be represented as a directed graph. We first mathematically formulate the problem as a Binary Quadratic Program (BQP), accounting for both computation and communication costs. We show that the problem is NP-hard. We then relax the problem as a Semi-Definite Program (SDP) and utilize a randomized rounding technique based on sampling from a suitably-formulated multi-variate Gaussian distribution. Furthermore, we derive the expected value of bottleneck time. Finally, we apply our proposed scheme on gossip-based federated learning as an application of iterative processes. Through numerical evaluations on the MNIST and CIFAR-10 datasets, we show that our proposed approach outperforms well-known scheduling techniques from distributed computing. In particular, for arbitrary settings, we show that it reduces bottleneck time by $91\%$ compared to HEFT and $84\%$ compared to throughput HEFT.
\end{abstract}

\begin{IEEEkeywords}
Bottleneck Time, Distributed Iterative Process, SDP, Randomized Rounding, Task Scheduling, Federated-Learning.
\end{IEEEkeywords}}

\maketitle

\IEEEdisplaynontitleabstractindextext

% \fi
\IEEEpeerreviewmaketitle

\IEEEraisesectionheading{\section{Introduction}\label{sec:introduction}}

% for IEEE Computer Society journal papers produced under \LaTeX\ using
% IEEEtran.cls version 1.8b and later.
\IEEEPARstart{F}{or} the emerging wave of applications such as Internet-of-Things (IoT) and mobile-data, training Machine Learning (ML) models may need to be performed in a
distributed fashion for reasons such as data privacy\footnote{Not allowing one centralized entity have access to data from many sources.}. 
This has given rise to Federated Learning (FL) frameworks which aim at preserving data privacy.  
Another reason for training ML models in a distributed manner is due to massive computations of their growing scale\footnote{For instance, the computations for deep neural networks remarkably increases as number of layers and hidden nodes increases.}\cite{massive-computation-ML-arxiv}-\cite{journal-massive-computation}, hence careful allocation of processing ML models on distributed and networked computers plays a crucial role in significantly reducing the execution time
\footnote{ 
High-performance ML training with privacy guarantees can be achieved on trusted distributed computing platforms by utilizing recently developed techniques~\cite{trusted_envi}.}.

Distributed ML applications such as FL, where model parameters are exchanged after certain number of iterations, fall under the umbrella of distributed iterative processes \cite{distributed-iterative-algo-Bertsekas}. Distributed iterative processes consist of multiple tasks with a given inter-task data dependency structure, i.e. each task generates inputs for certain other tasks.
Such a distributed iterative process can be described by a directed graph, known as task graph, where vertices represent tasks of the process and edges indicate the inter-task data dependencies.

In each iteration of executing a distributed iterative process, every task requires to be executed and its processed data needs to be transferred to computing resources where its successive tasks located on.  
The total time taken by the task with a dominant combined computational time (for executing a task) and communication time (to transfer the processed data of a task), which is referred to as bottleneck time, will be equal to the required time for an iteration. 
Since the total time required to execute an iterative process for a certain number of iterations is equal to the summation of time required to complete each iteration, 
minimizing bottleneck time would consequently lead to decreasing the completion time of the entire process.

% Therefore, minimizing bottleneck time would consequently lead to decreasing the completion time of the entire process.

Bottleneck time minimization can be achieved through efficient \emph{task scheduling} where tasks of an iterative process are assigned to appropriate distributed computing resources to be executed. 
Most prior task scheduling schemes are tailored to a particular class of task graphs called Directed Acyclic Graph (DAG) \cite{DAG-task-scheduling-Topcuoglu},\cite{HEFT},\cite{DAG-task-graph-genetic},\cite{DAG-task-graph-Sinnen}, while the task graphs of distributed iterative processes that we consider belong to a broader class of directed graphs (with or without cycles).      
Furthermore, significant number of existing task scheduling schemes (e.g., \cite{HEFT},\cite{makespan-TSUNGCHYAN},\cite{makespan-Johannes},\cite{makespan-MIN})  have focused on minimizing makespan, i.e. the time it takes to finish the execution for one set of inputs, which is meaningful only for a DAG-based task graph. Only few works have investigated minimizing bottleneck time (or equivalently maximizing throughput) such as \cite{HEFT-TP}.  

The underlying methodology for task scheduling can be categorized into heuristic-based algorithms (e.g. \cite{heuristic_new_Eskandari}- \cite{new_heuristic_journal_Pham}), meta-heuristic ones
(e.g. \cite{Kennedy-optimization},\cite{ADDYA-Simulatedannealing}-\cite{Fan-Simulated-Annealing}, \cite{SHISHID-Genetic-based},\cite{DASGUPTA-Genetic},\cite{genetic_Izadkhah}), and optimization-based schemes (e.g. \cite{Azar-convex-unrelated}-\cite{Skutella-SDP-no-comm}). 
%As far as the base of task scheduling schemes is concerned, most of prior work have studied heuristic-based algorithms for task scheduling \cite{heuristic_new_Eskandari}- \cite{new_heuristic_journal_Pham} and only small portion of them are meta-heuristic. 
One of the most well-known heuristic task scheduling scheme is HEFT \cite{HEFT} which we will consider as one of our benchmarks.
% Although heuristic schemes are used to be fast in finding emay lead to a solution far from the optimum, an efficient meta-heuristic scheme can guarantee a solution near the optimum. 
Although heuristic schemes are used to be fast and often get stuck in a local optimum, meta-heuristic and optimization-based schemes have gained significant attention as they are practically able to obtain a solution near the optimum \cite{survey-meta-nearoptimal},\cite{cite51-survey},\cite{cite52-survey}. 
Since task scheduling belongs to the class of NP-hard problems in its nature \cite{survey-meta-nearoptimal} \footnote{Task scheduling is a well-known NP-hard problem because of the solution space and required time to obtain the optimal solution.}, an appropriate relaxation of the optimization problem such as a Convex-based one can potentially lead to an efficacious performance \cite{convex-relaxation-is-good}.
Nevertheless, most Convex-based task scheduling schemes (e.g. \cite{Azar-convex-unrelated},\cite{Skutella-SDP-no-comm}) have focused on the case where there is no inter-task data dependency and the communication delay across distributed resources is negligible compared to computational time.  
However, there are reasons to explore other parts of the design space as processors today are remarkably quicker in computing and distributed processing may potentially operate across a wireless network with low bandwidth or wide-area network with high delays.

To our best knowledge, there is no prior work that has proposed a Semi-Definite Programming (SDP) relaxation for distributed iterative processes such that both computation and communication aspects are taken into account. 

The main contributions of this paper are as follows:
\begin{itemize}
    \item We formulate task scheduling of distributed iterative processes, which only requires task graph to a directed one, on distributed computing machines as an optimization problem.
    \item We propose a concrete SDP approximation for the aforementioned optimization problem and utilize a randomized rounding technique to achieve a feasible solution.
    \item We analyze the expected value of the bottleneck time of our proposed scheme.
    \item We provide a mathematical upper bound on the optimal solution. 
    \item We evaluate the performance of our proposed scheme on real data and Gossip-based federated learning and show that it outperforms HEFT \cite{HEFT} and another approach \cite{HEFT-TP} which investigated maximizing the throughput. 
\end{itemize}

\subsection{Prior Work}
Efficient task scheduling plays a crucial role in improving the utilization of computing resources as well as reducing required time to executing tasks.
Task scheduling can be categorized into multiple groups from different perspectives. For instance, from the type of tasks need to be processed perspective, task scheduling is traditionally divided into two categories, namely static and dynamic scheduling. The former is applicable when the information about tasks (such as required computational amount or deadline) and computing resources (such as processing power or communication delay) is available in advance while in the latter one the information of new task revealed during the execution of ongoing tasks. Another classification for task scheduling schemes is to divide them into two classes based on relationship of tasks: independent tasks and dependent tasks (realistic applications). Dependent tasks can be represented through a directed graph where tasks and inter-task data dependencies are represented as nodes and edges of the graph, respectively. The directed graph enforces inter-task dependencies by letting a task to be executed once all its predecessors are finished. 

The other way of categorizing task scheduling schemes has to do with the type of algorithms aiming at assigning tasks to compute resources. Heuristic, meta-heuristic, and optimization-based are three categories of task scheduling schemes. Heuristic task scheduling schemes can be divided into quite a few sub categories based on their objectives such as load balancing \cite{Ren-loadbalance-2012},\cite{Bhatia-loadbalance},\cite{KUMAR-loadbalance}, priority-based scheduling \cite{SUDARSAN-priority},\cite{Dubey-priority},\cite{list_scheduling}, , task  duplication \cite{duplication}, and clustering \cite{clustering}.

Since heuristic algorithms may considerably deviate from the optimal task scheduling, meta-heuristic and optimization-based schemes which aim at approximating the NP-hard optimization of task scheduling have attracted significant attention. Not only meta-heuristic and optimization-based schemes are suitable for solving large scale problems, they are practically efficient in leading to near optimal solutions. Some examples of meta-heuristic schemes are as follows: Particle Swarm Optimization \cite{Kennedy-optimization},  which , Simulated Annealing \cite{ADDYA-Simulatedannealing}-\cite{Fan-Simulated-Annealing}, Genetic-based approach \cite{SHISHID-Genetic-based},\cite{DASGUPTA-Genetic},\cite{genetic_Izadkhah}. 

Although there are very few work that have focused on providing a convex-relaxation based solution for the scheduling problem, they are not applicable to any distributed iterative process as they do not consider a general directed task graph (which may have cycles) \cite{Azar-convex-unrelated}-\cite{Skutella-SDP-no-comm}. Furthermore, these schemes have focused on different objective functions with different constraints (such as letting task split across distributed computing machines) than minimizing bottleneck time. Finally, these works have not considered the fact that different links/paths between distributed compute machines may experience different communication bandwidth. To our knowledge, this is the first work that a) considers general directed graphs, b) minimizes the bottleneck time taking into account both the compute costs and heterogeneous network communication costs, and c) provides a convex-relaxation based solution. 

The remaining of the paper is organized as follows: In the next section, we elaborate upon the problem formulation. In section \ref{sec:SDP}, the proposed SDP approximation of our problem is derived. Finally, in section \ref{sec:NR}, we show the numerical results of the performance of our proposed scheme against well-known approaches.

\section{Problem Statement}
\label{sec:PF}

In this section, we now formulate minimizing the bottleneck time of 
any distributed iterative process where it has inter-task dependency. We first briefly elaborate upon gossip-based federated learning as an example of distributed iterative processes, then we focus on expressing the problem formulation.

\subsection{Gossip-based Federated Learning}
To preserve data privacy in distributed ML such as FL, secure communication links can be established across trusted distributed computing resources\footnote{Through utilizing recently developed techniques such as ~\cite{trusted_envi}.}. 
A gossip-based FL scheme can be modeled as a network topology with a set of users denoted by $\mathcal U$ where each user $i$, $\forall i\in \mathcal U$, gossips its local model parameters to a pre-defined set of other users, denoted by ${\mathcal U}_i$. Then, the network topology can be represented via a directed graph  $G_{FL}:=(V_{FL},E_{FL})$ where $V_{FL}:=\{i|i\in {\mathcal U}\}$ and $E_{FL}:=\{e_{i,j}^{(FL)}|i\in {\mathcal U}, j\in {\mathcal U}_i\}$ indicate the set of vertices and edges, respectively. Each user aggregates the model parameters gossiped to it from other users and updates its local model parameters. Training convergence is accomplished by repeating the aforementioned procedure \cite{gossip-FL-segmented}.  

\subsection{Problem Formulation}
\begin{figure}[h]
\centering
\includegraphics[trim= 20mm 20mm 20mm 10mm,scale=.43]{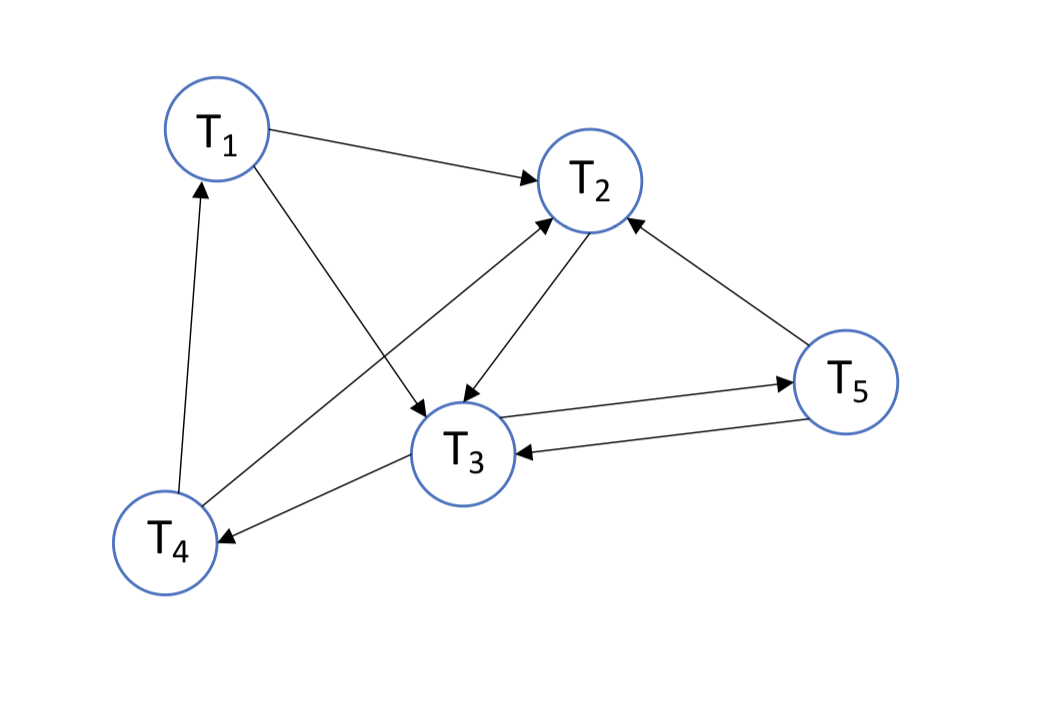}
\caption{An illustration of a task graph with five tasks.}
\label{fig:task_graph}
\end{figure}

We next show how to formally express bottleneck time minimization as an optimization problem. Every distributed iterative process (such as gossip-based FL) can be executed on a distributed computing platform where resources are interconnected via communication links. Hence, executing a distributed iterative process on distributed computing resources can be described by two separate directed graphs, namely \emph{task graph} and \emph{compute graph} which stand for the task structure of the distributed iterative process and the distributed computing platform, respectively. We next explain about each of these graphs as follows:

\noindent{\bf Task Graph:} Since there are dependencies across different tasks, meaning that a task generates inputs for certain other tasks, we can model this dependency through a \emph{directional} graph as depicted in Fig. \ref{fig:task_graph}. Unlike most prior work which considered Directed Acyclic Graph (DAG), we assume our task graph to be a general directed graph.  
Let us suppose to have $N_T$ tasks $\{T_i\}_{i=1}^{N_T}$ with a given task graph $G_{Task}:=(V_{Task},E_{Task})$ where $V_{Task}:=\{T_i\}_{i=1}^{N_T}$ and $E_{Task}:=\{e_{i,i'}\}_{(i,i')\in \Omega}$ respectively represent the set of vertices and edges (task dependencies) with $\Omega :=\{(i,i')|$ if task $T_i$ generates inputs for task $T_{i'}$ $\}$. Let us define ${\bf p}:= [p_1,\dots,p_{N_T}]^T$ as the required amount of computations of tasks.

\begin{figure}[h]
\centering
\includegraphics[trim= 10mm 10mm 10mm 10mm,scale=.43]{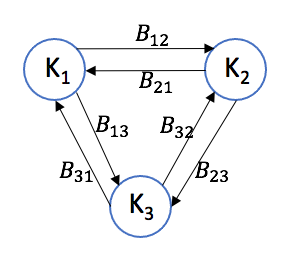}
\caption{An illustration of a compute graph with three machines and corresponding bandwidths shown on edges.}
\label{fig:compute_graph}
\end{figure}

\noindent{\bf Compute Graph:} Each task is required to be executed on a \emph{Compute} node (machine) which is connected to other compute nodes (machines) through communication links (compute node and machine are interchangeably used in this paper). Let us suppose to have $N_K$ compute nodes $\{K_j\}_{j=1}^{N_K}$. Regarding the execution speed of compute nodes, we consider vector ${\bf e}:= [e_1,\dots,e_{N_K}]^T$ as the executing speed of machines. 
The communication link delay between any two compute nodes can be characterized by bandwidth. In case of two machines not being connected to each other, we can assume the corresponding bandwidth is zero (infinite time for communication delay). Hence, the communication aspect of distributed computing nodes can be presented as a \emph{complete}\footnote{A complete graph is a type of graph in which any two different vertices are connected.} graph $G_{Compute}:=(V_{Compute},E_{Compute})$ where $V_{Compute}:=\{K_j\}_{j=1}^{N_K}$ and $E_{Compute}:=\{\frac{1}{b_{j,j'}}\}_{\forall j\neq j'}$ respectively indicate the set of compute nodes and links connecting them with $b_{j,j'}$ as the bandwidth of the link from machine $K_j$ to machine $K_{j'}$. An illustration of a compute graph with $N_K=3$ is shown in Fig. \ref{fig:compute_graph}.
%Please note that this definition supports the case of two machines not being connected (or having zero bandwidth) by assigning infinite time as communication delay between them.
Since the result of executing a task is a model with same number of parameters, we can assume the communication delay across machines is denoted by matrix $C\in \mathbb{R}^{N_K\times N_K}$.

We now elaborate upon how to formulate the bottleneck time of distributed iterative processes. In particular, we next present the objective function as well as constraints imposed due the nature of the problem.   

\noindent{\bf Objective Function:} we aim to obtain the optimal task \emph{mapper} function, denoted as $m(.):V_{Task} {\rightarrow} V_{Compute}$, to assign task $i$ $\forall i$ to machine $m(i)$ such that the bottleneck time is minimized. 
% The aforementioned task mapper function can be expressed as matrix $M\in \{0,1\}^{T\times K}$ where $[M]_{i,j}$ (component of $M$ at row $i$ and column $j$) is one if task $i$ is assigned to machine $j$;otherwise it is zero. 
Regarding the bottleneck time, it is referred to maximum \emph{compute-communicate} time over all tasks for a given task mapper matrix $M\in \{0,1\}^{N_T\times N_K}$ (equivalent of the mapper function $m(.)$), where $[M]_{i,j}=1$ if task $i$ is assigned to compute machine $j$; otherwise $[M]_{i,j}=0$. By defining $S:= (G_{Task},G_{Compute},m(V_{Task}))$, the compute-communicate time of task $i$ can be expressed as follows

\begin{equation}\label{compute-and-communicate}
\begin{aligned} 
t_{comp-comm}^{(i)}(S) := t_{comp}^{(i)}(S) + t_{comm}^{(i)}(S)~~\forall i, 
\end{aligned}    
\end{equation}
where $t_{comp}^{(i)}(S)$ is the required time for task $i$ to be executed on machine $m(i)$ (compute time) and $t_{comm}^{(i)}(S)$ is the time for the result to be transmitted to machines which are specified to run the immediate successive tasks of task $i$ (communicate time). 
Therefore, the bottleneck time would be
\begin{equation}\label{bottleneck}
\begin{aligned} 
t_{Bottleneck}(S):= \max_{i}{~~~~t_{comp-comm}^{(i)}(S)}. \end{aligned}    
\end{equation}

The objective function can be formally written as:
\begin{equation}\label{objective}
\begin{aligned} 
t_{Bottleneck}^*:= \min_{M\in\{0,1\}^{N_T\times N_K}}{\max_{i}{~~~~t_{comp-comm}^{(i)}(S)}}. 
\end{aligned}
\end{equation}

\noindent{\bf Constraints:} since each task needs to be executed on a machine and then its results to be sent to others machines where executing the successive tasks, we can write these constraints as follows
\begin{equation}\label{constraints}
\begin{aligned} 
\sum_{j}{[M]_{i,j}}= 1 ~~~~~~\forall i, 
\end{aligned}
\end{equation}
or equivalently rewriting as 
\begin{equation}\label{constraints_multiplication}
\begin{aligned} 
M{\bf 1}_{N_K\times 1}= {\bf 1}_{N_T\times 1}. 
\end{aligned}
\end{equation}

\noindent{\bf Optimization Problem:} by considering above objective function and constraints, the optimization problem would be as 
\begin{equation}\label{optimization}
\begin{aligned} 
\min_{M\in\{0,1\}^{N_T\times N_K}}{\max_{i}{~~~~t_{comp-comm}^{(i)}(S)}}
\\\text{s.t.  }M{\bf 1}_{N_K\times 1}= {\bf 1}_{N_T\times 1}.
\end{aligned}
\end{equation}

We next write the objective function in terms of required task processing vector ${\bf p}$, machine execution speeds vector ${\bf e}$, communication delay matrix $C$, and task mapper function in closed form.   
Each term of (\ref{compute-and-communicate}) can be further derived as 

\begin{itemize}
    \item {$t_{comp}^{(i)}(S)$:} Since each machine runs all tasks assigned to it in parallel\footnote{The CPU allocation is proportional to the size of required computations for tasks.}, the required time to execute task $i$ is 
    \begin{equation}\label{compute-time}
    \begin{aligned} 
    t_{comp}^{(i)}(S)&= \frac{\sum_{r:m(r)=m(i)}{p_r}}{e_{m(i)}}
    \overset{(a)}{=}\sum_{\ell=1}^{N_K}{\frac{{\bf p}^TM_{\ell}}{M_i{\bf e}}[M]_{i,\ell}}~~~~~~~ \forall i, 
    \end{aligned}    
    \end{equation}
    where $M_i$ indicates the $i$th row of matrix $M$ and $(a)$ follows from the fact that each row of matrix $M$ has single 1 (due to the imposed constraint). By further simplification, (\ref{compute-time}) can be rewritten as follows
    \begin{equation}\label{compute-time_simplified}
    \begin{aligned} 
    t_{comp}^{(i)}(S)&=M_iDM^T{\bf p} = I_iMDM^T{\bf p} ~~~~~~~ \forall i, 
    \end{aligned}    
    \end{equation}
    where $I_i$ denotes an indicator row-vector of size $N_T$ with $i$-th element equals 1 and the remaining are zeros,
    %$I_i:= [0,\dots,0,\underbrace{1}_{i\text{th component}},0,\dots,0]$, 
    $D:= diag(1 \varoslash{\bf e})$ and $\varoslash$ denotes component-wise division. By defining $m:= vec(M)$ and the fact $\text{trace}\{AXBX^T\}=vec(X)^T(B^T\otimes A)vec(X)$ ($\otimes$ indicates Kronecker product), we can rewrite (\ref{compute-time_simplified}) as 
    \begin{equation}\label{compute-time_simplified_vec}
    \begin{aligned} 
    t_{comp}^{(i)}(S)&=\text{trace}\{I_iMDM^T{\bf p}\}=\text{trace}\{MDM^T{\bf p}I_i\}
    =m^T(D^T\otimes {\bf p}I_i)m ~~~~~~~ \forall i, 
    \end{aligned}    
    \end{equation}
    
    \item {$t_{comm}^{(i)}(S)$:} The result of computed task $i$ needs to be transmitted to all machines assigned to execute successive tasks $i'$ where $e_{i,i'} \in E_{Task}$. Considering constraints (\ref{constraints}), the communication delay time for sending result of task $i$ from machine $m(i)$ to machine $i'$ is $[C]_{m(i),m(i')}$. 
    Since the result is required to be sent to all machines running the successive tasks, we would have 
    \begin{equation}\label{comm-time}
    \begin{aligned} 
    t_{comm}^{(i)}(S)&=\max_{i':e_{i,i'}\in E_{Task}}{[C]_{m(i),m(i')}} ~~~~~~~ \forall i. 
    \end{aligned}    
    \end{equation}
    Further simplification leads to
    \begin{equation}\label{comm-time_rewrite}
    \begin{aligned} 
    t_{comm}^{(i)}(S)&=\max_{i':e_{i,i'}\in E_{Task}}{M_iCM_{i'}^T}
    \\&=\max_{i':e_{i,i'}\in E_{Task}}{I_iMCM^TI_{i'}^T}
    \\&=\max_{i':e_{i,i'}\in E_{Task}}{m^T(C^T\otimes I_i^TI_{i'})m}~~~~~~~ \forall i,
    \end{aligned}    
    \end{equation}

\end{itemize}

By combining (\ref{compute-time_simplified_vec}) and (\ref{comm-time_rewrite}), the objective function can be written as follows

\begin{equation}\label{optimization_vec}
\begin{aligned} 
\min_{m\in\{0,1\}^{N_TN_K\times 1}}{\max_{i,i':e_{i,i'}\in E_{Task}}{\{m^TQ_{i,i'}m\}}}
\\\text{s.t.  }Hm= {\bf 1}_{N_T\times 1},
\end{aligned}
\end{equation}
where 
\begin{equation}\label{optimization_vec_definition}
\begin{aligned} 
Q_{i,i'}&:= D^T\otimes {\bf p}I_i+C^T\otimes I_i^TI_{i'}
\\H &:= {\bf 1}_{1\times N_K} \otimes I_{N_T\times N_T}
\end{aligned}
\end{equation}
and $I_{N_T\times N_T}$ is identical matrix of size $N_T$ by $N_T$.

Optimization problem (\ref{optimization_vec}) can be rewritten as 
\begin{equation}\label{optimization_vec_t}
\begin{aligned} 
\min_{m\in\{0,1\}^{N_TN_K\times 1},t}{t}
\\\text{s.t.  }m^TQ_{i,i'}m &\leq t ~~~~~~\forall i,i':e_{i,i'}\in E_{Task},
\\Hm&= {\bf 1}_{N_T\times 1}.
\end{aligned}
\end{equation}

Since components of vector $m$ in (\ref{optimization_vec_t}) only take integer values of 0 or 1, (\ref{optimization_vec_t}) is not Convex, hence obtaining the optimal solution for this BQP is cumbersome.

\noindent{\bf Remark 1:} After making matrices $Q_{i,i'}$ symmetric, i.e. replacing $Q_{i,i'}$ with $\frac{Q_{i,i'}+Q_{i,i'}^T}{2}$, (\ref{optimization_vec_t}) is not necessarily Semi-Definite Positive (or Semi-Definite Negative).   

Before proceeding with the relaxation of (\ref{optimization_vec_t}), let us first focus on two special cases of (\ref{optimization_vec_t}) in the following theorem and proposition.

\noindent{\bf Theorem 1:} \textit{In case of assuming the communication delay is negligible compared to computational time, i.e. $C={\bf 0}_{N_K\times N_K}$, no inter-task data dependency, and allowing at most single task to be executed on each machine, the optimal task mapper function $m(.)$ can be obtained by assigning the available task with highest required computation to the available machine with the fastest execution speed , after sorting tasks and machines in terms of their required computations and execution speeds, respectively.}

\noindent{\bf \textit{Proof}:} By sorting machines in term of their execution speeds (meaning the first machine is the fastest) and tasks in term of the required computations (meaning the first task needs the highest amount of computations), then task at index $\ell$ of the sorted is executed by machine at index $\ell$ of the sorted machines. Let us assume tasks $i$ and $k$ where $p_i\geq p_k$ (i.e. task $i$ has more computations than task $k$) are respectively executed on machines $j$ and $j'$. Then the bottleneck time would be $t_1=\max\{t_{others},\frac{p_i}{e_j},\frac{p_k}{e_{j'}} \}$ where $t_{others}$ represents the time for completing other tasks. By swapping task assignment of tasks $i$ and $k$, i.e. tasks $i$ and $k$ are respectively executed on machines $j'$ and $j$, then the required time would be $t_2=\max\{t_{others},\frac{p_k}{e_j},\frac{p_i}{e_{j'}} \}\leq t_1$ due to $\frac{p_i}{e_{j'}}\leq \frac{p_k}{e_j} \leq \max\{\frac{p_i}{e_{j'}}, \frac{p_k}{e_j}\}$ and $\frac{p_k}{e_{j}}\leq \frac{p_i}{e_{j'}} \leq \max\{\frac{p_k}{e_{j}}, \frac{p_i}{e_{j'}}\}$. Therefore, one can easily see the aforementioned assignment of sorted tasks to sorted machines leads to an optimal solution.

\noindent{\bf Proposition 1:} Suppose the communication delay to be negligible compared to computational time, i.e. $C={\bf 0}_{N_K\times N_K}$, no inter-task data dependency, and the execution speed of all compute nodes to be identical, then the optimization problem (\ref{optimization_vec_t}) would be expressed as follows 
\begin{equation}\label{optimization_vec_t_subsetsum}
\begin{aligned} 
\min_{m(.)}{\max_{i}{\sum_{\ell:m(\ell)=m(i)}{p_{\ell}}}},
\end{aligned}
\end{equation}
which is exactly the same as the problem of load balancing (minimizing the maximum load\footnote{The load of each machine is defined as the sum of computational processing amounts of tasks assigned to it.} across $N_K$ machines). In other words, (\ref{optimization_vec_t_subsetsum}) aims at assigning all $N_T$ tasks across $N_K$ machines such that the total computation time of all machines are \emph{nearly the same}. The problem is NP-complete due to the followings: 
\begin{itemize}
    \item A non-deterministic polynomial-time algorithm can solve (\ref{optimization_vec_t_subsetsum}) by guessing an assignment of tasks into $N_K$ machines, then verifying in polynomial time if all machines have the same computational load . 
    \item The reduction from well-known NP-complete problem of Set Partitioning into our problem works in polynomial time. In particular, given $N_T$, $N_K$, $\{p_i\}_{i=1}^{N_T}$ and a target value  $\theta=\frac{{\sum_i}{p_i}}{N_K}$, if there is a solver to our problem (i.e. verifying in polynomial time whether there exists an assignment of tasks with work load of at most $\Theta$ for all machines), then the solver can determine if there is a solution to Set Partition problem with $\{p_i\}_{i=1}^{N_T}$ as the instance inputs of Set Partition problem.  
\end{itemize}

%%%%%%%%%%%%%%%%%%%%%%%%%%%%%%%%%%%%%%%%%%%%%%%%%%%%%%
%%%%%%%%%%%%%%%%%%%%% Section IV %%%%%%%%%%%%%%%%%%%%%
%%%%%%%%%%%%%%%%%%%%%%%%%%%%%%%%%%%%%%%%%%%%%%%%%%%%%%
\section{Semi-Definite Programming (SDP) Relaxation}
\label{sec:SDP}

Due to difficulty in solving (\ref{optimization_vec_t}), we next elaborate upon how we can relax the problem to a SDP problem which is easier to solve while leading to a desirable solution.
%resulting in close solution to the original problem.

Since it is easier to apply approximation (on \emph{homogenized} quadratic programming) of (\ref{optimization_vec_t}) when $m\in \{-1,+1\}^{N_TN_K\times 1}$ rather than $m\in \{0,1\}^{N_TN_K\times 1}$, we rewrite (\ref{optimization_vec_t}) as

\begin{equation}\label{optimization_vec_t_plus_minus}
\begin{aligned} 
\min_{{\bf x}\in\{-1,1\}^{N_TN_K\times 1},t}&{t}
\\\text{s.t.  }{\bf x}^TQ_{i,i'}{\bf x}+2\Big({\bf 1}_{N_TN_K\times 1}^TQ_{i,i'}{\bf x}\Big)+&{\bf 1}_{N_TN_K\times 1}^TQ_{i,i'}{\bf 1}_{N_TN_K\times 1} \leq 4t ~~~~~~\forall i,i':e_{i,i'}\in E_{Task},
\\H{\bf x}= (2-N_K)&{\bf 1}_{N_T\times 1},
\end{aligned}
\end{equation}
by replacing ${\bf m}$ with $\frac{{\bf x}+{\bf 1}}{2}$. We can reformulate (\ref{optimization_vec_t_plus_minus}) as the following optimization problem:
\begin{equation}\label{optimization_vec_t_plus_minus_X}
\begin{aligned} 
\min_{{\bf x}\in \mathbb{R}^{N_TN_K\times 1}, X\in \mathbb{R}^{N_TN_K \times N_TN_K},t}&{t}
\\\text{s.t.  }<Q_{i,i'},X>+2\Big({\bf 1}_{N_TN_K\times 1}^TQ_{i,i'}{\bf x}\Big)+&{\bf 1}_{N_TN_K\times 1}^TQ_{i,i'}{\bf 1}_{N_TN_K\times 1} \leq 4t ~~~~~~\forall i,i':e_{i,i'}\in E_{Task},
\\H{\bf x}&= (2-N_K){\bf 1}_{N_T\times 1},
\\X&={\bf x}{\bf x}^T,
\\\text{diag}(X)&=1,
\end{aligned}
\end{equation}
where $<Q_{i,i'},X> :=\text{trace}\{Q_{i,i'}X\}$. The optimization problem (\ref{optimization_vec_t_plus_minus_X}) is not Convex due to constraint $X={\bf x}{\bf x}^T$. A well-known SDP technique is to replace constraint $X={\bf x}{\bf x}^T$ with $X \succeq {\bf x}{\bf x}^T$ where $\succeq$ indicate semi-definite positive used for matrices. Therefore, the relaxed version of (\ref{optimization_vec_t_plus_minus_X}) would be 
\begin{equation}\label{optimization_vec_t_plus_minus_X_relaxed_cons}
\begin{aligned} 
\min_{{\bf x}\in \mathbb{R}^{N_TN_K\times 1}, X\in \mathbb{R}^{N_TN_K \times N_TN_K},t}&{t}
\\\text{s.t.  }<Q_{i,i'},X>+2\Big({\bf 1}_{N_TN_K\times 1}^T&Q_{i,i'}{\bf x}\Big)
+{\bf 1}_{N_TN_K\times 1}^TQ_{i,i'}{\bf 1}_{N_TN_K\times 1} \leq 4t ~~~~~~\forall i,i':e_{i,i'}\in E_{Task},
\\H{\bf x}&= (2-N_K){\bf 1}_{N_T\times 1},
% \end{aligned}
% \end{equation}
% \begin{equation}\nonumber
% \begin{aligned} 
\\\begin{bmatrix}
   X &
   {\bf x} \\
   {\bf x}^T &
   1 
   \end{bmatrix} &\succeq 0,
\\\text{diag}(X)&=1.
\end{aligned}
\end{equation}
Due to non-homogeneous structure of (\ref{optimization_vec_t_plus_minus_X_relaxed_cons}), i.e. appearance of both $X$ and ${\bf x}$ in quadratic constraints which causes difficulty in rounding the final solution to a feasible point, we aim at re-formulating (\ref{optimization_vec_t_plus_minus_X_relaxed_cons}) into a new homogenized optimization problem as follows   

\begin{equation}\label{optimization_vec_t_plus_minus_X_relaxed_cons_z}
\begin{aligned} 
\min_{{\bf x}\in\mathbb{R}^{N_TN_K\times 1}, X\in \mathbb{R}^{N_TN_K \times N_TN_K},u\in\mathbb{R},t}&{t}
\\\text{s.t.  }<Q_{i,i'},X>+2u{\bf 1}_{N_TN_K\times 1}^T&Q_{i,i'}{\bf x}+u^2{\bf 1}_{N_TN_K\times 1}^TQ_{i,i'}{\bf 1}_{N_TN_K\times 1} \leq 4t ~\forall i,i':e_{i,i'}\in E_{Task},
\\uH{\bf x}&= (2-N_K){\bf 1}_{N_T\times 1},
\\\begin{bmatrix}
   X &
   {\bf x} \\
   {\bf x}^T &
   1 
   \end{bmatrix} &\succeq 0,
\\\text{diag}(X)&=1,
\\u^2 &= 1.
\end{aligned}
\end{equation}

By defining ${\bf {\tilde x}} := [{\bf x}^T,u]^T$, we can rewrite (\ref{optimization_vec_t_plus_minus_X_relaxed_cons_z}) as follows

\begin{equation}\label{optimization_vec_t_plus_minus_X_new}
\begin{aligned} 
&\min_{{\bf {\tilde x}}\in \mathbb{R}^{(N_TN_K+1)\times 1}, {\tilde X} \in \mathbb{R}^{(N_TN_K+1) \times (N_TN_K+1)},t}{t}
\\&~~~\text{s.t.  }<{\tilde Q}_{i,i'},{\tilde X}> \leq 4t ~~\forall i,i':e_{i,i'}\in E_{Task},
\\&~~~~~~~~~~~~<A_i,{\tilde X}>= 0 ~~~~~~\forall i\in\{1,\dots,N_T\},
\\&~~~~~~~~~~~~~~\begin{bmatrix}
   {\tilde X} &
   {\bf {\tilde x}} \\
   {\bf {\tilde x}}^T &
   1 
   \end{bmatrix} \succeq 0,
\\&~~~~~~~~~~~~~~~~\text{diag}({\tilde X})=1,
\end{aligned}
\end{equation}
where

\begin{equation}\label{optimization_vec_t_plus_minus_X_new_definitions}
\begin{aligned} 
{\tilde Q}_{i,i'} &:=
\begin{bmatrix}
   Q_{i,i'} &
   \frac{Q_{i,i'}{\bf 1}_{N_TN_K}}{2} \\
   \frac{{\bf 1}_{N_TN_K}^TQ_{i,i'}}{2} &
   {\bf 1}_{N_TN_K}^TQ_{i,i'}{\bf 1}_{N_TN_K}
   \end{bmatrix} ~\substack{\forall i,i':\\e_{i,i'}\in E_{Task}}
\\A_i &:= 
\begin{bmatrix}
   {\bf 0}_{(N_TN_K+1)\times(N_TN_K+1)} &
    \frac{H_i^T}{2}\\
   \frac{H_i}{2} &
   (N_K-2)
   \end{bmatrix}~\substack{\forall i\in \\ \{1,\dots,N_T\}}.
\end{aligned}
\end{equation}

To round the solution of (\ref{optimization_vec_t_plus_minus_X_new}), denoted by ${\tilde X}^*$ and ${\bf {\tilde x}^*}$, we apply a similar randomized technique as \cite{Boyd_docu}. In particular, we first collect samples ${\bf z}\sim \mathcal{N}({\bf 0},{{\tilde X}^{*}})$ and take $sign({\bf z})$ to have binary values of $-1$ or $+1$. Then, we keep the samples which satisfy the constraints. To prevent failure in finding such points holding all constraints, we can alter constraints $H{\bf x}= (2-N_K){\bf 1}_{N_T\times 1}$ to $H{\bf x}\geq (2-N_K){\bf 1}_{N_T\times 1}$ without changing the optimal value \footnote{Replacing equality constraint with inequality for this constraint means that allowing tasks to be executed more than once which clearly, at the optimal point, results in the same solution as the case allowing tasks to be executed at only one machine. To see this, one can drop all but one of the machines each task needs to be run on, then it is clear these two cases leads to the same optimal solution}. Finally, we select the sample point with the lowest objective value. 
We next derive the expected value of the bottleneck time of our randomized scheme, then we elaborate upon an upper bound on the bottleneck time of the optimal solution.
%In the next part, we elaborate upon an upper bound of bottleneck time for our randomized scheme. 
%Since our proposed algorithm picks the sample point which has the minimum bottleneck time, the average bottleneck time would be an upper bound for our randomized method.

\subsection{Expected Value Analysis}
In this section, we provide the average bottleneck time of our proposed technique for a given task graph $G_{Task}$ and compute graph $G_{Compute}$. By defining ${\bf \hat z}:=sign({\bf z})$, the expected value of the bottleneck time of our proposed scheme is
\begin{equation}\label{expected}
\begin{aligned} 
\max_{i,i':e_{i,i'}\in E_{Task}} \frac{1}{4}\mathbb{E}_{\bf z}[{\bf \hat z}^T{\tilde Q}_{i,i'}{\bf \hat z}].
\end{aligned}
\end{equation}

To obtain above expected value, we need to find $\mathbb{E}[{\bf \hat z}^T{Q}{\bf \hat z}]$ where $Q:={\tilde Q}_{i,i'}$ for given $i$ and $i'$, as follows
\begin{equation}\label{expected_derivation}
\begin{aligned} 
\mathbb{E}_{\bf z}[{\bf \hat z}^TQ{\bf \hat z}]
&\overset{(a)}{=}\frac{2}{\pi}\sum_{w,v}{[Q]_{w,v}\arcsin([{\tilde X}^{*}]_{w,v})}
\end{aligned}
\end{equation}
where $(a)$ follows from the proof presented in Appendix A. %\ref{appendix}. 

\subsection{Upper Bound on the Optimal Solution}

Since (\ref{optimization_vec_t_plus_minus_X_new}) is the relaxed version of (\ref{optimization_vec_t}), it is clear the optimal solution to (\ref{optimization_vec_t}), denoted as $OPT$, is greater than or equal to the solution of the SDP problem (\ref{optimization_vec_t_plus_minus_X_new}), i.e.
\begin{equation}\label{optimization_expected_lower_bound}
\begin{aligned} 
\max_{\substack{i,i':\\e_{i,i'}\in E_{Task}}} \frac{1}{4}\sum_{w,v}{[Q_{i,i'}]_{w,v}[{\tilde X}^{*}]_{w,v}}\leq OPT.
\end{aligned}
\end{equation}

On the other hand, the optimal solution to (\ref{optimization_vec_t}) is less than or equal to any other feasible solution including the solution to the minimization of the expected value of bottleneck time expressed as follows
\begin{equation}\label{optimization_expected}
\begin{aligned} 
\min_{t}&{~~t}
\\\text{s.t.   }\mathbb{E}_{\bf z}[{\bf \hat z}^T{\tilde Q}_{i,i'}{\bf \hat z}]& \leq 4t ~~\forall i,i':e_{i,i'}\in E_{Task}.
\end{aligned}
\end{equation}

Since (\ref{expected_derivation}) can be upper-bounded as follows
\begin{equation}\label{expected_derivation_upperbound}
\begin{aligned} 
&\frac{2}{\pi}\sum_{w,v}{[Q]_{w,v}\arcsin([{\tilde X}^{*}]_{w,v})}
\overset{(b)}{\leq}\sum_{w,v}{[Q]_{w,v}(0.112+0.878[{\tilde X}^{*}]_{w,v})}
\end{aligned}
\end{equation}
where $(b)$ follows from and the fact $\frac{2}{\pi}\arcsin(x)\leq 0.112+0.878x$ $\forall |x|\leq1$, one can easily see 
\begin{equation}\label{optimization_expected_upper_bound}
\begin{aligned} 
OPT \leq \max_{\substack{i,i':\\e_{i,i'}\in E_{Task}}} \sum_{w,v}{[Q_{i,i'}]_{w,v}(0.112+0.878[{\tilde X}^{*}]_{w,v})}
\end{aligned}
\end{equation}

We are guaranteed to reach this expected value after sampling sufficient (feasible) points ${\bf \hat z}$, hence we know that the optimal value is between the solution of the SDP and the solution to (\ref{optimization_expected}).

\section{Numerical Results}
\label{sec:NR}
In this section, we provide the numerical results obtained by applying our scheme in comparison with well-known techniques utilized for distributed computing such as HEFT \cite{HEFT} and throughput HEFT \cite{HEFT-TP}. As far as the simulation settings are concerned, we perform scheduling of tasks for two different scenarios, 1) \emph{arbitrary distributed iterative process with pre-defined settings}\footnote{The required amount of computations for tasks and execution speed of computing machines are known in advance. This case is utilized to illustrate the performance of different schedulers under any arbitrary settings.} and 2) gossip-based federated learning for the sake of classification of MNIST and CIFAR-10 datasets. 

\subsection{Distributed Iterative Process with Pre-defined Settings}
In this part, we provide the numerical results for arbitrary task computation vector ${\bf p}$ and arbitrary execution speed vector ${\bf e}$ of distributed machines.    
Since HEFT-based schemes require the task graph to be a Directed Acyclic Graph (DAG), we next present how to construct a new DAG from a given task graph in order to feed into HEFT-based algorithms. 

\begin{figure*}[t]
\centering
\includegraphics[trim= 60mm 70mm 120mm 30mm,scale=.7]{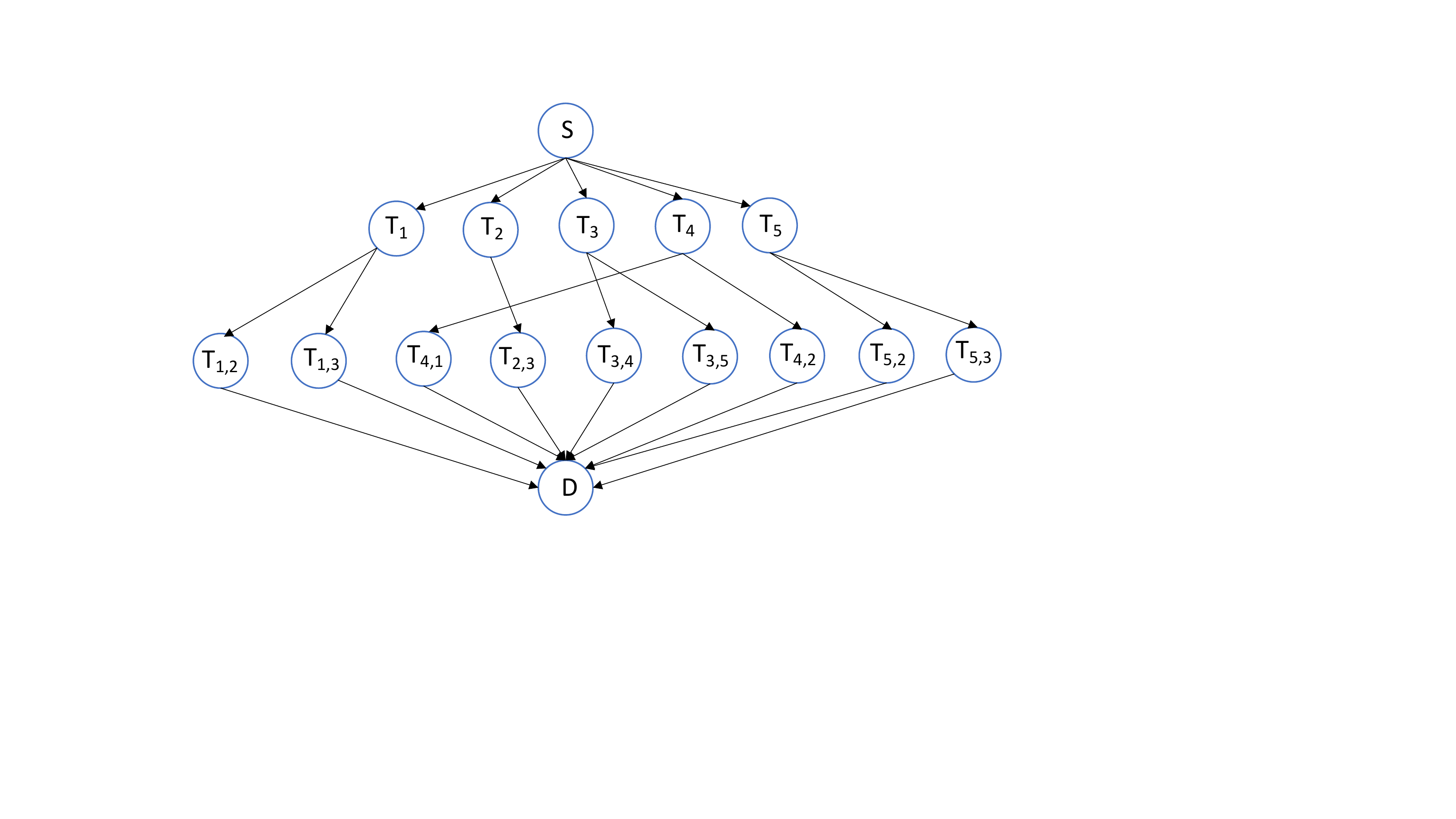}
\caption{The corresponding DAG of the task graph $G_{Task}$ depicted in Fig. \ref{fig:task_graph} to be fed into HEFT-based schemes.}
\label{fig:entire}
\end{figure*}
\subsubsection{Creating a new DAG from $G_{Task}$ for HEFT-based Schemes}
Let us define $G_{DAG}:=(V_{DAG},E_{DAG})$ as the corresponding DAG of the task graph $G_{Task}$. We determine set of vertices $V_{DAG}$ first, then set of edges $E_{DAG}$. Regarding $V_{DAG}$, it consists of all vertices of task graph $G_{Task}$ as well as the following vertices:

\begin{itemize}
    \item Source vertex $S$. 
    \item Intermediate vertices $T_{i,j}$'s for all $i$ and $j$ such that $e_{ij}\in E_{Task}$ (i.e. task $T_i$ is the parent of task $T_j$). 
    \item Destination vertex $D$.
\end{itemize}

As far as $E_{DAG}$ is concerned, it includes the set of edges of task graph $G_{Task}$ and the following edges:
\begin{itemize}
    \item Outgoing edges of vertex $S$: set of edges $\{e_{S,T_i}\}_{i:T_i \in V_{Task}}$ which connects $S$ to vertex $T_i$ for all $T_i \in V_{Task}$.
    \item Incoming edges of intermediate vertices $T_{i,j}$'s: set of edges $\{e_{T_i,T_{i,j}}\}$ which connects vertex $T_i$ to vertex $T_{i,j}$ for all $i$ and $j$ such that task $T_i$ is the parent of task $T_j$ (i.e. $e_{ij}\in E_{Task}$). 
    \item Incoming edges of vertex $D$: 
    set of edges $\{e_{T_{i,j},D}\}$ which connects vertex $T_{i,j}$ to vertex $D$ for all $i$ and $j$.
\end{itemize}

Therefore, we can formally present the aforementioned DAG as $G_{DAG}=(V_{DAG},E_{DAG})$ where $V_{DAG}:=V_{Task}\cup\{S,\{T_{i,j}\}_{i,j:e_{i,j}\in E_{Task}},D\}$ and $E_{DAG}:=E_{Task} \cup \{e_{S,T_i}\}_{i:T_i\in V_{Task}}\cup \{e_{T_i,T_{i,j}}\}_{i,j:e_{i,j}\in E_{Task}} \cup \{e_{T_{i,j},D}\}_{i,j:e_{i,j}\in E_{Task}}$. An illustration of the corresponding DAG of the task graph $G_{Task}$ of Fig. \ref{fig:task_graph} is shown in Fig. \ref{fig:entire}.

\subsubsection{Numerical Results for Pre-defined Settings}
\begin{figure}[h]
\centering
\includegraphics[trim= 20mm 10mm 10mm 10mm,scale=.29]{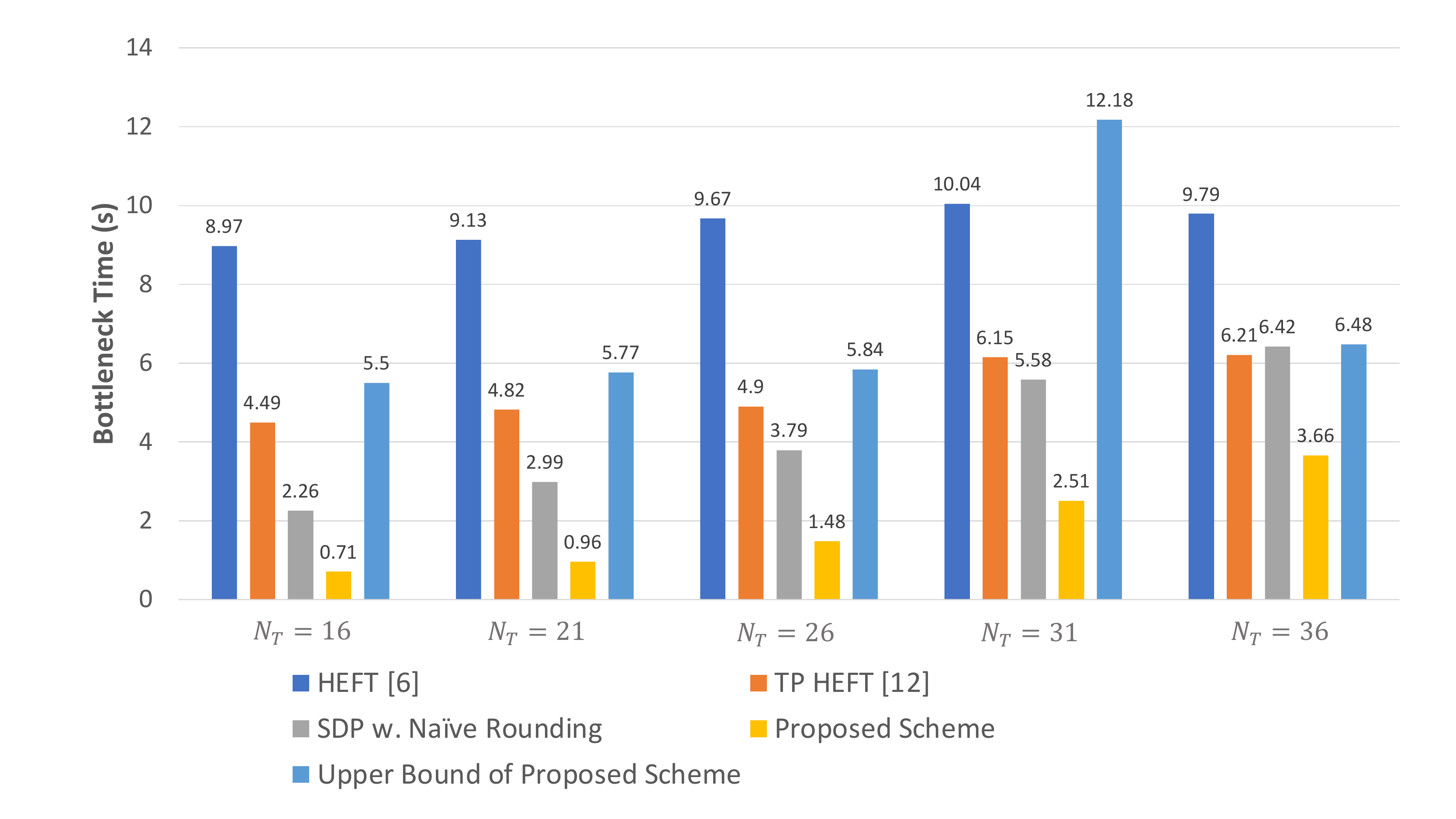}
\caption{Bottleneck time of different schemes, namely HEFT \cite{HEFT}, Throughput (TP) HEFT \cite{HEFT-TP}, SDP method with naive rounding, our proposed scheme, and the upper bound of our proposed approach, for different number of tasks.}
\label{fig:comparison_tasks}
\end{figure}

Figure \ref{fig:comparison_tasks} shows the bottleneck time of our proposed scheme along SDP method with naive rounding (i.e. rounding the solution of SDP to the closest integer) against HEFT \cite{HEFT} and Throughput (TP) HEFT \cite{HEFT-TP} for the following setting: $N_K=4$ (four compute nodes), components of communication matrix $C$ are i.i.d. and drawn from $\mathcal N(0,\sqrt{10})$, execution speed of compute nodes are i.i.d and drawn from $\mathcal N(0,\sqrt{15})$, required computation of tasks are i.i.d. and drawn from $\mathcal N(0,1)$. 
%The task graph is also randomly generated such that the minimum and maximum degrees of its nodes are 6 and 7, respectively. 
As it can be seen, our proposed scheme outperforms HEFT \cite{HEFT} due to the fact that HEFT schedule tasks based on the average communication delay of links while our scheme schedules based on actual communication delay through solving an optimization problem. In particular, our proposed scheme leads to 63\%-91\% reduction in bottleneck time compared to HEFT \cite{HEFT} and 41\%-84\% compared to TP HEFT \cite{HEFT-TP}. 
Fig. \ref{fig:comparison_tasks} further shows the upper bound for our proposed scheme. One can see that even the upper bound of our proposed scheme is considerably lower than HEFT in most cases.
% \begin{figure}[h]
% \centering
% \includegraphics[trim= 90mm 10mm 90mm 0mm,scale=.46]{images/picture1.png}
% \caption{The corresponding DAG of the task graph $G_{Task}$ depicted in Fig. \ref{fig:task_graph} to be fed into HEFT-based schemes.}
% \label{fig:comparison_tasks}
% \end{figure}

\begin{figure}[h]
\centering
\includegraphics[trim= 10mm 10mm 10mm 10mm,scale=.29]{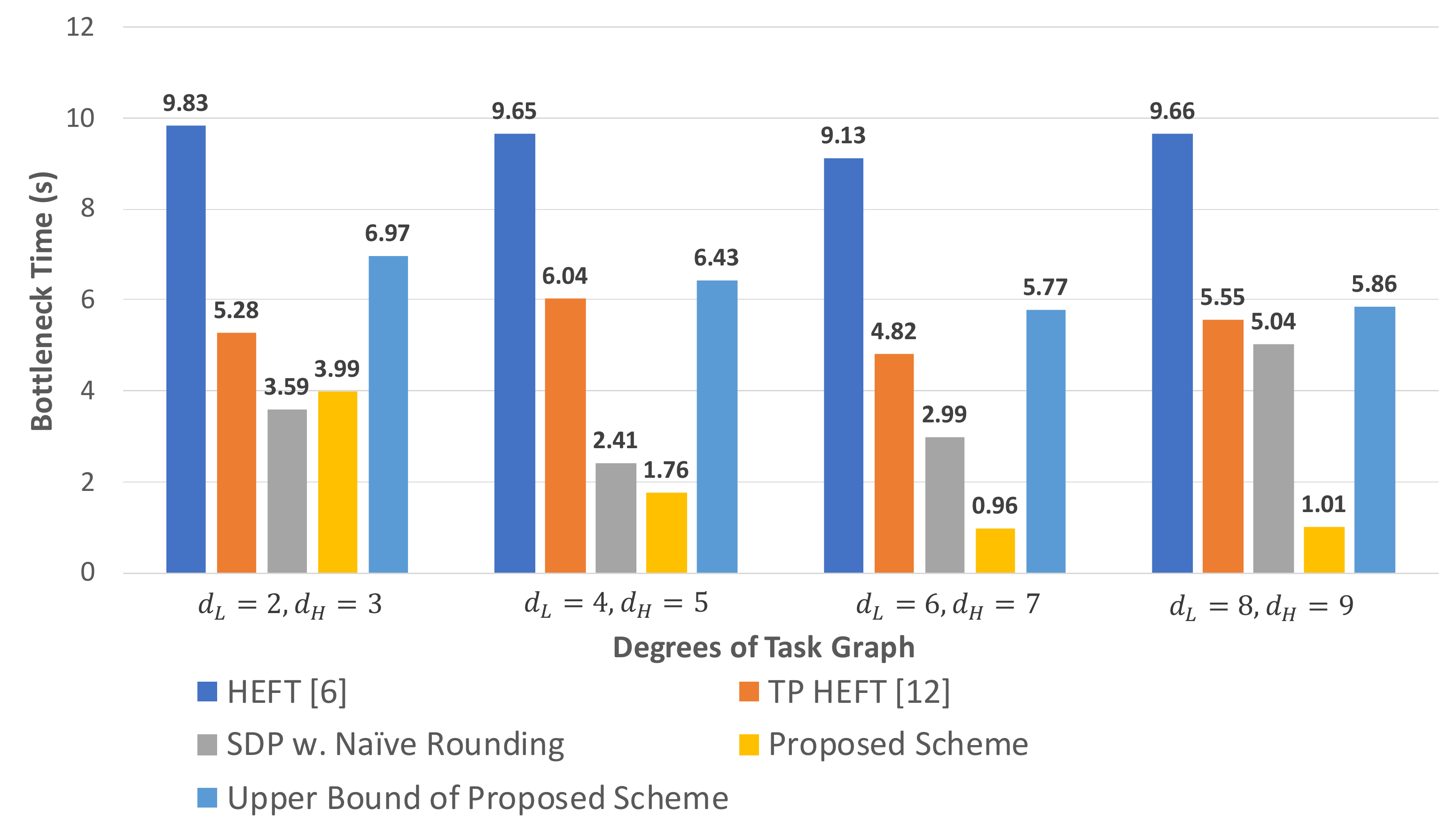}
\caption{The comparison of bottleneck time of the following schemes: HEFT \cite{HEFT}, Throughput (TP) HEFT \cite{HEFT-TP}, SDP method with naive rounding, our proposed scheme, and the upper bound of our proposed approach, for different degrees of the task graphs.}
\label{fig:comparison_tasks_degree}
\end{figure}

Figure \ref{fig:comparison_tasks_degree} illustrates the comparison of our proposed scheme as well as the SDP approach with naive rounding against well-known schemes such as HEFT \cite{HEFT} and  TP HEFT \cite{HEFT-TP} in terms of bottleneck time for different setting of degrees of vertices of task graphs. In Fig.  \ref{fig:comparison_tasks_degree}, $d_L$ and $d_H$ represent the minimum and the maximum degree of vertices of task graphs, respectively. Regarding the remaining settings, we consider the same settings as before with $N_T=21$. The first observation of Fig. \ref{fig:comparison_tasks_degree} is that the upper bound of our proposed scheme is considerably lower than HEFT scheme \cite{HEFT} (around  29\%-39\%). The other interesting observation is that our proposed scheme significantly outperforms HEFT \cite{HEFT} and TP HEFT \cite{HEFT-TP} (59\%-90\% reduction in bottleneck time compared to HEFT \cite{HEFT} and 25\%-82\% compared to TP HEFT \cite{HEFT-TP}) as task graphs become more dense (degree of vertices becomes larger). The reason behind this gain is due to the fact that each task has more successors (children tasks) in dense task graphs. Hence, it is possible for one of these children to be scheduled on a machine with poor communication in HEFT-based schemes. Our scheme takes care of all communication links while HEFT-based schemes only consider average communication for links in their task mapping phase. Therefore, for a HEFT-based scheme, it is plausible to make inefficient assignment by 
scheduling a task on a machine with poor communication, hence resulting in large bottleneck time.

\begin{figure}[h]
\centering
\includegraphics[scale=.52]{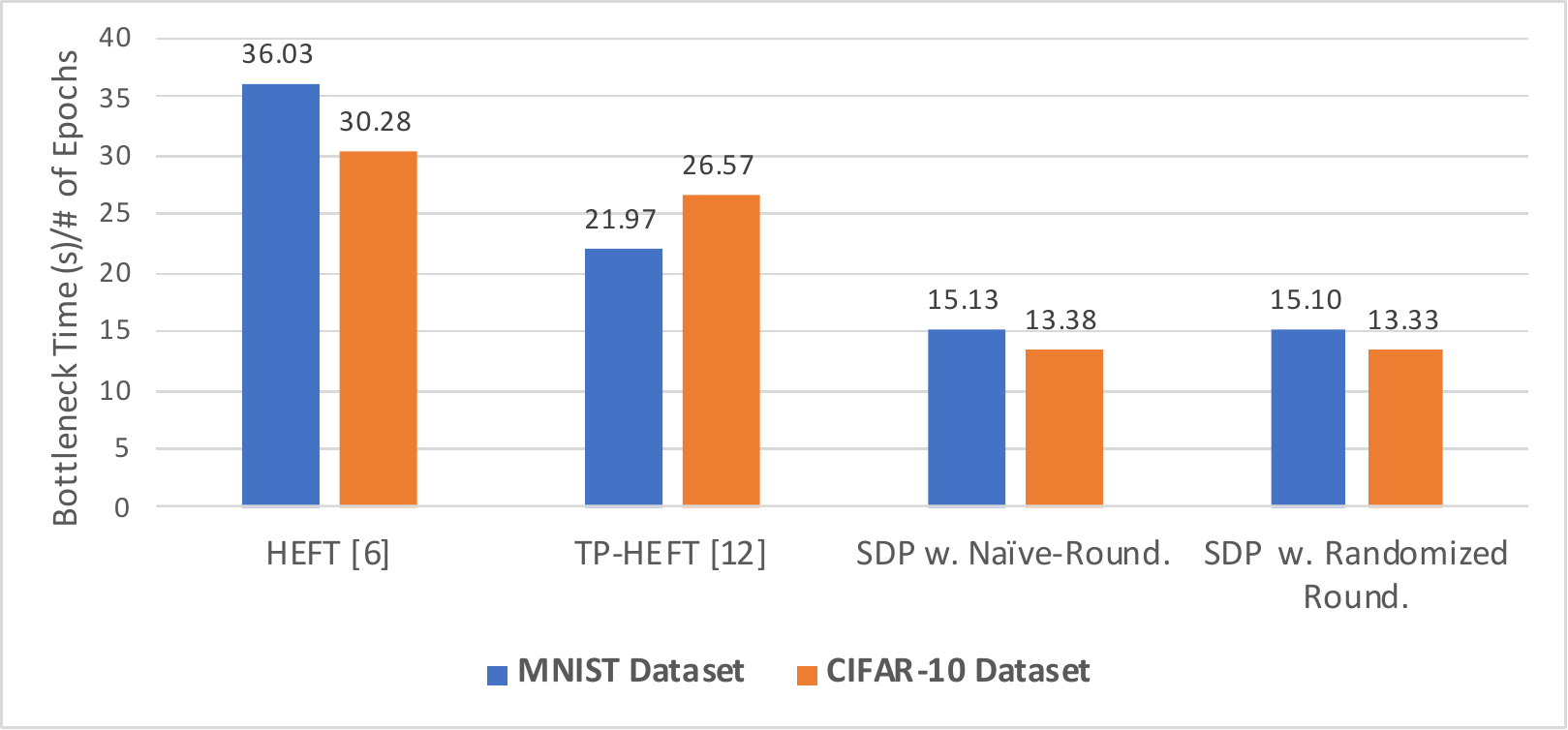}
\caption{Bottleneck time of executing gossip-based federated learning of $N_T=10$ tasks on $N_K=4$ distributed machines for four different schemes, namely HEFT \cite{HEFT}, Throughput (TP) HEFT \cite{HEFT-TP}, SDP method with naive rounding, and our proposed scheme (SDP with randomized rounding).}
\label{fig:FL_simulation}
\end{figure}

\subsection{Gossip-based Federated Learning}
In this part, we investigate the bottleneck time of performing an application of our optimization problem (\ref{optimization}). In particular, the gossip-based federated learning can be formulated as optimization problem (\ref{optimization}) where each task is associated with part of whole dataset (each task can be viewed as an \emph{user} in a real-world gossip-based federated learning problem). 
In order to simulate the gossip-based federated learning, we considered 10 tasks that form a random task graph of $G_{task}$ where the degree of each vertex is random and drawn from $\text{Unif}(6,7)$ distribution.
As far as the gossiping of model parameters is concerned, all users first partition their associated data into $w$ \emph{chunks} and perform training a neural network model with a chunk of data, then they send their obtained models to pre-determined set of tasks(or users)\footnote{The dependancy of tasks are enforced by the task graph.}. Upon receiving model parameters, each user aggregates\footnote{A simple aggregation is a weighted average of the model parameters.} the models, updates its local model and begins training process with its next chunk of data. Once each user uses its entire dataset (going over all its chunks of data) during training its model, it starts the training with a new epoch.

We select the classification of MNIST and CIFAR-10 datasets,  through applying Convolutional Neural Network (CNN), as two examples of gossip-based federated learning which runs on distributed computing machines. Regarding the CNN incorporated in our simulation, we considered a CNN model with two convolutional layers as well as three fully connected layers. 
Based on the network delay and the size of model parameters that need to be gossiped across machines, we consider each component of communication matrix $C$, i.e. $C_{i,j}$(for $i\neq j$), to be random and drawn from $\text{Unif}(0,1)$ distribution.

Fig. \ref{fig:FL_simulation} shows the bottleneck time of running the gossip-based federated learning on distributed computing systems with four different schedulers, namely HEFT \cite{HEFT}, TP-HEFT \cite{HEFT-TP}, our SDP approach with naive rounding, and our proposed SDP scheme with randomized rounding. 
Since all schedulers assign tasks to compute machines based on required processing vector ${\bf p}$, we design a \emph{pilot} phase\footnote{This step is performed before running the gossip-based federated learning.} to estimate tasks' required computations amount. 
To do so, since we evenly divide dataset among tasks, all tasks required computations are the same. Furthermore, for simplicity, we assume all compute machines are homogeneous, i.e. having the same execution speed. Hence, each task consider a small portion of its data to be used for the pilot step. To determine ${\bf p}$, we first measure the time required to train the model for a task on a compute machine with pilot data, then multiplying it with execution speed of the compute machine. As one can easily observe, our two proposed SDP-based approach outperform HEFT \cite{HEFT} and TP-HEFT \cite{HEFT-TP} in terms of bottleneck time of gossip-based federated learning while reaching high accuracy.

\section{Conclusion}
\label{sec:conc}
We proposed a new task scheduling scheme so as to speed up iterative processes which are run on distributed computing resources. We mathematically formulated our task scheduling problem as a BQP, then provided a Semi-Definite Programming based approximation to our problem as well as utilizing a randomized rounding technique. Moreover, we analyzed the the expected value of bottleneck time and derived an upper bound for the optimal BQP. Finally, as a concrete application example, we considered gossip-based federated learning which fits  distributed iterative process. We showed that our proposed scheme outperforms well-known techniques such as \cite{HEFT} and \cite{HEFT-TP}.

% or


%%%%%%%%%%%%%%%%%%%%%%%%%%%%%%%%%%%%%%%%%%%%%%%%%%%%%%%%%%%%
%%%%%%%%%%%%%%%%%%%%%%%%%%%%%%%%%%%%%%%%%%%%%%%%%%%%%%%%%%%%

% \section{}

\ifCLASSOPTIONcompsoc
  \section*{Acknowledgments}
\else
  \section*{Acknowledgment}
\fi
This material is based upon work supported by Defense Advanced Research Projects Agency (DARPA) under Contract No. HR001117C0053. Any views, opinions, and/or findings expressed are those of the author(s) and should not be interpreted as representing the official views or policies of the Department of Defense or the U.S. Government.

\ifCLASSOPTIONcaptionsoff
  \newpage
\fi



\bibliographystyle{IEEEtran}
\bibliography{main}
\appendices
\section{}\label{appendix}
In this section, we prove that $\mathbb{E}_{\bf z}[{\bf \hat z}^TQ{\bf \hat z}]=\frac{2}{\pi}\sum_{w,v}{[Q]_{w,v}\arcsin([\Sigma]_{w,v})}$ for ${\bf z}\sim \mathcal{N}({\bf 0},\Sigma)$.
\begin{equation}\label{arcsin}
\begin{aligned} 
&\mathbb{E}_{\bf z}[{\bf \hat z}^TQ{\bf \hat z}]
=\mathbb{E}[\text{trace}\{Q{\bf \hat z }{\bf \hat z }^T\}]  
=\mathbb{E}[\sum_{w,v}{[Q]_{w,v}{\hat z}_w{\hat z}_v}]
\\&=\sum_{w,v}{[Q]_{w,v}\mathbb{E}[{\hat z}_w{\hat z}_v]}
% \end{aligned}
% \end{equation}
% \begin{equation}\nonumber
% \begin{aligned} 
\\&=\sum_{w,v}{[Q]_{w,v}\mathbb{E}[sign({z}_w)sign({z}_v)]}
\\&=\sum_{w,v}{[Q]_{w,v}\Big({Pr}[{z}_w\geq 0,{z}_v\geq 0]+
{Pr}[{z}_w\leq 0,{z}_v\leq 0]}
-{Pr}[{z}_w\leq 0,{z}_v\geq 0]-
{Pr}[{z}_w\geq 0,{z}_v\leq 0]\Big),
\end{aligned}
\end{equation}
where $Pr[A]$ denotes the probability of event $A$.
By defining random variable $z:=\frac{z_v-\rho z_w}{\sqrt{1-\rho^2}}$ where $\rho:=cov(z_w,z_v)$, one can easily verify that $z\perp z_w$ with $z$ and $z_w$ have zero-mean unit-variance normal distribution. Considering this, we have
\begin{equation}\label{arcsin}
\begin{aligned} 
&{Pr}[{z}_w\geq 0,{z}_v\geq 0]={Pr}[{z}_w\geq 0,{z}\geq \frac{\rho}{\sqrt{1-\rho^2}}z_w]
\\&=\int_{z_w=0}^{\infty}{ \int_{z=az_w}^{\infty}{ 
\frac{1}{\sqrt{2\pi}}e^{-\frac{z_w^2}{2}}
.\frac{1}{\sqrt{2\pi}}e^{-\frac{z^2}{2}}dzdz_w
}  }
\\&=\frac{1}{2\pi}(\frac{\pi}{2}-\arctan(a))=
\frac{1}{2\pi}(\frac{\pi}{2}+\arcsin(\rho)),
\end{aligned}
\end{equation}
where $a=\frac{-\rho}{\sqrt{1-\rho^2}}$. By following similar approach for ${Pr}[{z}_w\leq 0,{z}_v\leq 0]
$, ${Pr}[{z}_w\leq 0,{z}_v\geq 0]$, and ${Pr}[{z}_w\geq 0,{z}_v\leq 0]$, we can simplify $(\ref{arcsin})$ as 
\begin{equation}\label{arcsin_final}
\begin{aligned} 
&\mathbb{E}_{\bf z}[{\bf \hat z}^TQ{\bf \hat z}]
=\frac{2}{\pi}\sum_{w,v}{[Q]_{w,v}\arcsin([\Sigma]_{w,v})}.
\end{aligned}
\end{equation}
%%%%%%%%%%%%%%%%%%%%%%%%%%%%%%%%%%%%%%%%%%%%

\end{document}